\documentclass[aps,prl,showpacs,twocolumn,floats,psfig]{revtex4}
\usepackage{amssymb}
\usepackage{amsbsy}
\usepackage{amsmath}
\usepackage{epsfig}

\newcommand\bea{\begin{eqnarray}}
\newcommand\eea{\end{eqnarray}}
\newcommand\beq{\begin{equation}}
\newcommand\eeq{\end{equation}}

\newcommand{\non}{\nonumber}
\newcommand{\bib}{\bibitem}
\newcommand{\al}{\alpha}
\newcommand{\de}{\delta}
\newcommand{\lam}{\lambda}
\newcommand{\om}{\omega}

\newcommand{\da}{\dagger}
\newcommand{\la}{\langle}
\newcommand{\ra}{\rangle}
\newcommand{\vk}{\vec k}

\begin{document}

\title{Defect production in non-linear quench across a quantum critical point}
\author{Diptiman Sen$^1$, K. Sengupta$^2$, and Shreyoshi Mondal$^2$}
\affiliation{$^1$ Center for High Energy Physics, Indian Institute of Science,
Bangalore, 560 012, India \\ $^2$ TCMP division, Saha Institute of Nuclear 
Physics, 1/AF Bidhannagar, Kolkata 700 064, India}

\date{\today}

\begin{abstract}
We show that the defect density $n$, for a slow non-linear power-law quench
with a rate $\tau^{-1}$ and an exponent $\al>0$, which takes the system
through a critical point characterized by correlation length and dynamical
critical exponents $\nu$ and $z$, scales as $n \sim \tau^{-\al \nu d/
(\al z\nu+1)}$ [$n \sim (\al g^{(\al-1)/\al}/\tau)^{\nu d/(z\nu+1)}$]
if the quench takes the system across the critical point at time $t=0$
[$t=t_0 \ne 0$], where $g$ is a non-universal constant and $d$ is the system
dimension. These scaling laws constitute the first theoretical results for
defect production in non-linear quenches across quantum critical points and
reproduce their well-known counterpart for a linear quench ($\al=1$) as a
special case. We supplement our results with numerical studies of well-known
models and suggest experiments to test our theory.
\end{abstract}


\maketitle

Quantum phase transitions have been studied extensively for several
years \cite{sachdev1}. Such transitions are accompanied by diverging
length and time scales \cite{sachdev1} leading to the absence of
adiabaticity close to the quantum critical point (QCP). Thus a time
evolution of a parameter $\lam(t)=\lam_0 |t/\tau|^\al {\rm sign}(t)$
in the system Hamiltonian, characterized by a rate $1/\tau$ and an
exponent $\al>0$, which takes such a system across a QCP 
located at $\lam = \lam_c$, leads to a failure of the
system to follow the instantaneous ground state and hence to the
production of defects \cite{kz1,bd1}. All the previous studies on
such systems have been restricted to linear quenches ($\al=1$)
\cite{ks1,dziar1,cardy1,levitov1,das1,sen1,bd2,caneva1,wz1}. For such
quenches, the density of these defects $n \sim \tau^{-\nu d/(z\nu +1)}$,
where $\nu$ and $z$ are the correlation length and
the dynamical critical exponents characterizing the critical point,
and $d$ is the system dimension \cite{anatoly1,anatoly2}. On the
experimental side, trapped ultracold atoms in optical lattices
provide possibilities of realization of several model quantum spin
systems and are particularly suitable for studying their
non-equilibrium dynamics \cite{bloch1,duan}. Experimental studies of
defect production have indeed been undertaken for a spin-one Bose
condensate\cite{sk1}. Although these experiments can easily
investigate non-linear quenches, there have been no theoretical
studies of such quench phenomena so far.

In this letter, we show that a slow non-linear power-law quench, as
discussed above, through a QCP leads to a density
of defects which scales either as $n \sim \tau^{-\al \nu d/(\al z\nu
+1)}$ or as $n \sim (\al g^{(\al-1)/\al}/\tau)^{\nu d/(z\nu+1)}$
(where $g$ is a non-universal model dependent constant), depending
on whether the quench parameter $\lam$ vanishes or stays finite at
the critical point. Such a scaling law for the defect density
generalizes its earlier known counterpart for linear quenches
\cite{anatoly1} and thus constitutes a significant extension of our
understanding of quench dynamics across a QCP. Our
results, to the best of our knowledge, also constitute the first
theoretical investigation of defect production due to non-linear
power-law quenches. We supplement our theoretical results with
numerical studies of the one-dimensional Ising and Kitaev models,
and also suggest realistic experiments to test our theory.

We begin our analysis with a study of a model Hamiltonian in $d$
dimensions of the form
\bea H &=& \sum_k ~\psi^{\dagger} (\vec k) ~H(\vec k;t) 
~\psi(\vec k), \non \\
H (\vec k;t) &=& [\lambda(t) + b(\vec k)] \tau_3 + \Delta (\vec k) \tau_+ 
+ \Delta^{\ast} (\vec k) \tau_-, \label{ham1} \eea
where $\tau_{i=1,2,3}$ are the usual Pauli matrices with $\tau_{\pm}
=(\tau_1 \pm i \tau_2)/2$, $b(\vk)$ and $\Delta(\vk)$ are model
dependent functions, and $\psi(\vk)= (c'_1(\vk),c'_2(\vk))$ represents 
fermionic operators. Many of the model systems with a QCP 
characterized by $\nu=z=1$, such as the Ising and XY
models \cite{dziar1,levitov1} in $d=1$ and the extended Kitaev
models \cite{chen1,lee1,sen2} in $d=2$, can be mapped onto such a
fermionic Hamiltonian via standard Jordan-Wigner transformation. We
first consider the case where the system passes through a gapless
point at $t=0$ and $\vk=\vk_0$. Note that in this case both $b(\vk)$
and $\Delta(\vk)$ must vanish at $\vk=\vk_0$. In what follows, we
shall also assume that $|\Delta(\vk)| \sim |\vk -\vk_0|$ and $
b(\vk) \sim |\vk-\vk_0|^{z_1}$ at the critical point, where $z_1 \ge
1$ so that $E \sim |\vk-\vk_0|$ and $z=1$. In the rest of the analysis, we
set $\hbar=1$, and scale $t \to t\lam_0$, $\tau \to \tau \lam_0$,
$\Delta(\vk) \to \Delta(\vk)/\lam_0$, and $b(\vk) \to b(\vk)/\lam_0$.

The dynamics of the such a system is governed by the Schrodinger
equation given by $i \partial_t \psi(\vk) = H (\vk;t) \psi(\vk)$
which leads to the equation governing the time evolution of $c_1(\vk) =
c_1'(\vk) e^{i \int^t dt' [|t'/\tau|^{\al} {\rm sign}(t') + b(\vk)]}$,
\bea \left(\frac{d^2}{dt^2} + 2i \left[\left|\frac{t}{\tau}\right|^{\al}
{\rm sign}(t)+ b(\vk) \right] \frac{d}{dt} + |\Delta(\vk)|^2 \right)c_1 (\vk)
=0. \label{ceq1} \eea
Now we scale $t \to t \tau^{\al/(\al+1)}$ so that Eq. (\ref{ceq1}) becomes
\bea \left(\frac{d^2}{dt^2} + 2i \left[|t|^{\al}{\rm sign}(t) + b(\vk)
\tau^{\frac{\al}{\al+1}} \right] \frac{d}{dt} \right. \non \\
+ \left.|\Delta(\vk)|^2 \tau^{\frac{2\al}{\al+1}}\right) c_1(\vk)
=0. \label{ceq2} \eea Noting that the system was in the ground state
$(c_1(\vk),c_2(\vk))=(1,0)$ at the beginning of the quench at $t
=-\infty$, we find, using Eq. (\ref{ceq2}), that the defect
probability $p(\vk)$ must be given by \beq p(\vk) = \lim_{t \to \infty} 
|c_1(\vk ,t)|^2 = f\left[b(\vk) \tau^{\frac{\al}{\al+1}};|\Delta(\vk)|^2
\tau^{\frac{2\al}{\al+1}}\right], \eeq where $f$ is a function whose
analytical form is not known for $\al \ne 1$. Nevertheless, we note
that for a slow quench (large $\tau$), $p(\vk)$ becomes appreciable
only when the instantaneous energy gap $\de E = 2
[(\lam(t)+b(\vk))^2 +|\Delta(\vk)|^2]^{1/2}$ becomes small at some
point of time during the quench. Consequently, $f$ must vanish when
either of its arguments are large. Thus for a slow quench (large
$\tau$), the defect density $n = \int_{\rm BZ} d^d k/^d \,
f\left[b(\vk) \tau^{\frac{\al}{\al+1}};|\Delta(\vk)|^2
\tau^{\frac{2\al}{\al+1}}\right]$ (where ${\rm BZ}$ denotes the
Brillouin zone) receives its main contribution from values of $f$
near $\vk=\vk_0$ where both $b(\vk)$ and $\Delta(\vk)$ vanish and
can be written as (extending the range of momentum integration to
$\infty$) $n \simeq \int d^d k/(2\pi)^d \, f\left[|\vk-\vk_0|^{z_1}
\tau^{\frac{\al}{\al+1}};|\vk-\vk_0|^{2}
\tau^{\frac{2\al}{\al+1}}\right]$. Now scaling $\vk \to (\vk -
\vk_0) \tau^{\al/(\al+1)}$, we find that \bea n &=& \tau^{-
\frac{\al d}{\al+1}}\int \frac{d^d k}{(2\pi)^d}
f(|\vk|^{z_1} \tau^{\al(1-z_1)/(\al+1)};|\vk|) \non \\
& \simeq & \tau^{- \frac{\al d}{\al+1}} \int \frac{d^d k}{(2\pi)^d} 
f(0;|\vk|)~ \sim ~\tau^{- \frac{\al d}{\al+1}}, \label{sca1} \eea
where in arriving at the last line, we have used $z_1 > 1$ and $\tau \to
\infty$. (If $z_1 = 1$, the integral in the first line is independent of
$\tau$, so the scaling argument still holds). Note that for $\al=1$, Eq.
(\ref{sca1}) reduces to its counterpart for a linear quench \cite{anatoly1}.

Next we generalize our results for a critical point with arbitrary
$\nu$ and $z$. To this end, we consider a generic time dependent
Hamiltonian $H_1[t]\equiv H_1[\lam(t)]$, whose states are labeled by
$|\vk \ra$ and $|0\ra$ denotes the ground state. If there is a
second order phase transition, the basis states change continuously
with time during this evolution and can be written as $|\psi(t)\ra =
\sum_{\vk} a_{\vk}(t) |\vk[\lam(t)]\ra$ and the defect density can
be obtained in terms of the coefficients $a_{\vk} (t)$ as $n =
\sum_{\vk \ne 0} |a_{\vk}(t\to \infty)|^2$ so that one gets
\cite{anatoly1} \bea n \simeq \int \frac{d^d k}{(2 \pi)^d} \Big|
\int_{-\infty}^\infty d \lam \la \vk|\frac{d}{d \lam} |0 \ra e^{i
\tau \int^\lam d \lam' \de \om_{\vk} (\lam')} \Big|^2, \label{defect2} \eea
where $\de \om_{\vk} (\lam)=\om_{\vk} (\lam)-\om_0(\lam)$ are the 
instantaneous excitation energies, and we have replaced the sum over $\vk$ 
by a $d$-dimensional momentum integral. We note, following Ref.
\onlinecite{anatoly1}, that near a critical point, $\de
\om_{\vk} (\lam) = \Delta F(\Delta/|\vk|^z)$, where $\Delta$ is the
energy gap, $z$ is the dynamical critical exponent and $F(x) \sim
1/x$ for large $x$. Also, since the quench term vanishes at the
critical point, $\Delta \sim |\lam|^{\al z \nu}$ for a non-linear
quench, one can write $\de \om_{\vk} (\lam) = |\lam|^{\al z \nu}
F'(|\lam|^{\al z \nu}/|\vk|^z)$ where $F'(x) \sim 1/x$ for large
$x$. Further, one has $\la \vk|\frac{d}{d \Delta}|0\ra = |\vk|^{-z} G(\Delta
/|\vk|^z)$ near a critical point where $G(0)$ is a constant. This allows us
to write $\la \vk|\frac{d}{d \lam}|0\ra = \frac{\lam^{\al z \nu -1}}{|\vk|^z}
G'(\lam^{\al z \nu}/|\vk|^z)$ where $G'(0)$ is a constant
\cite{sachdev1,anatoly1}. Putting these in Eq. (\ref{defect2}) and changing
the integration variables to $\eta = \tau^{\al \nu/(\al z \nu + 1)} |\vk|$
and $ \xi = |\vk|^{-1/(\al \nu)} \lam$, we find that
\bea n ~\simeq ~C ~\tau^{-\al \nu d/(\al z \nu +1)}, \label{defect3} \eea
where $C$ is a non-universal number independent of $\tau$.

Next we focus on the case where the quench term does not vanish at the
QCP for $\vk = \vk_0$. We again consider the Hamiltonian
$H(\vk)$ in Eq. (\ref{ham1}), but now assume that the critical point
is reached at $t=t_0 \ne 0$. This renders our previous scaling argument
invalid since $\Delta (\vk_0) = 0$ but $b(\vk_0) \ne 0$. In this situation,
$|t_0/\tau| = g^{1/\al}$, where $g=|b(\vk_0)|$ is a non-universal model
dependent constant, so that the energy gap $\de E$ may vanish at the
critical point for $\vk = \vk_0$. We now note that the most important
contribution to the defect production comes from times near $t_0$ and
from momenta near $k_0$. Therefore, we expand the diagonal terms in
$H(\vk)$ about $t=t_0$ and $\vk =\vk_0$ to obtain
\bea H' &=& \sum_{\vk} \psi^{\da} (\vk) ~\Big[\left\{\al g^{(\al-1)/\al}
\left(\frac{t-t_0}{\tau}\right)+ b' (\de \vk)\right\} \tau_3 \non \\
&& ~~~~~~~~~~~~~~ +\Delta(\vk) \tau_+ + \Delta^{\ast}(\vk) 
\tau_- \Big] ~\psi(\vk), \label{ham2} \eea
where $b'(\de \vk)$ represents all terms in the expansion of
$b(\vk)$ about $\vk=\vk_0$ and we have neglected all terms $R_n =
(\al-n+1)(\al-n+2)...(\al-1) g^{(\al-n+1)/\al} |(t-t_0)/\tau|^{n}
{\rm sign}(t)/n!$ for $n>1$ in the expansion of $t/\tau$ about $t_0$. We
shall justify neglecting these higher order terms shortly.

Eq. (\ref{ham2}) describes a linear quench of the system with
$\tau_{\rm eff}(\al) = \tau/(\al g^{(\al-1)/\al})$. Hence one can
use the well-known results of Landau-Zener dynamics \cite{lz1} to
write an expression for the defect density $n = \int_{{\rm BZ}} d^dk
~p(\vk) = \int_{{\rm BZ}} d^d k \exp[ -\pi |\Delta(\vk)|^2 \tau_{\rm
eff}(\al)]$. For a slow quench, the contribution to $n$ comes from
$\vk$ near $\vk_0$ so that \bea n ~\sim ~\tau_{\rm eff}(\al)^{-d/2} ~=~ 
\left(\al g^{(\al-1)/\al}/\tau \right)^{d/2}. \label{defect4}
\eea Note that for $\al=1$, we get back the familiar result $n \sim
\tau^{-d/2}$ as a special case and the dependence of $n$ on the
non-universal constant $g$ vanishes. Also, since the quench is
effectively linear, one can use the results of Ref. \cite{anatoly1}
to find the scaling of the defect density when the critical point at
$t=t_0$ is characterized by arbitrary $\nu$ and $z$, 
\bea n ~\sim ~\left(\al g^{(\al-1)/\al}/\tau\right)^{\nu d/(z\nu +1)}.
\label{defect5} \eea

Next we justify neglecting higher order terms $R_n$. We note that
the significant contribution to $n$ comes at times $t$ when the
energy levels of $H'(\vk ;t)$ (Eq. (\ref{ham2})) for a given $\vk$ are
close to each other: $(t-t_0)/\tau \sim \Delta(\vk)$. Also, for a
slow quench, the contribution to the defect density is significant
only when $p(\vk)$ is significant, {\it ie.}, when $|\Delta(\vk)|^2
\sim 1/\tau_{\rm eff}(\al)$. Using these arguments, it is easy to
see that $R_n/R_{n-1} = (\al-n+1)g^{-1/\al}(t-t_0)/(n\tau) \sim
(\al-n+1)/(n \sqrt{\tau})$. Thus we find that all higher order
terms $R_{n>1}$, which were neglected in arriving at Eq. (\ref{defect4}), 
are unimportant in the limit of slow quench (large $\tau$).

\begin{figure}
\rotatebox{0}{\includegraphics*[width=\linewidth]{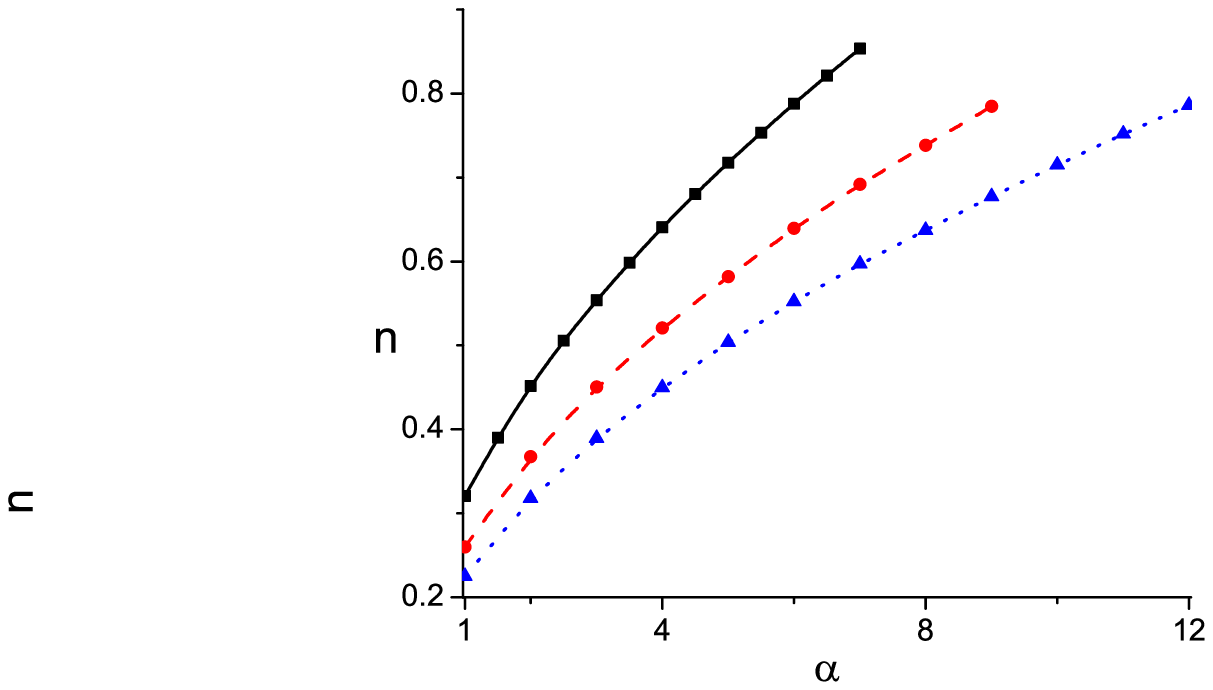}}
\caption{Variation of the defect density $n$ with the quench exponent $\al$ 
for representative values of $\tau=10$ (black solid line), $\tau=15$ (red 
dashed line) and $\tau=20$ (blue dotted line). A polynomial fit of the form 
$n = a \al^{b}$ yields exponents which are very close to the theoretical 
result $1/2$ for all values of $\tau$.} \label{fig1} \end{figure}

The scaling relations for the defect density $n$ given by Eqs.
(\ref{defect3}) and (\ref{defect5}) represent the central results of
this letter. For such power-law quenches, unlike their linear
counterpart, $n$ depends crucially on whether the quench term
vanishes at the critical point. For quenches which do not vanish at
the critical point, $n$ scales with the same exponent as that of a
linear quench, but is characterized by a modified non-universal
effective rate $\tau_{\rm eff}(\al)$. If, however, the quench term
itself vanishes at the critical point, we find that $n$ scales with
a novel $\al$ dependent exponent $ \al \nu d/(\al z \nu +1)$. For
$\al=1$, $\tau_{\rm eff}(\al) = \tau$ and $\al \nu d/(\al z \nu + 1)
= \nu d/(z\nu +1)$; hence both Eqs. (\ref{defect3}) and
(\ref{defect5}) reproduce the well-known defect production law for
linear quenches as a special case \cite{anatoly1}. We note that the
scaling of $n$ will show a cross-over between the expressions given
in Eqs. (\ref{defect3}) and (\ref{defect5}) near some value of $\tau
= \tau_0$ which can be found by equating these two expressions; this
yields $\tau_0 \sim |b (\vk_0)|^{- z \nu - 1/\al}$. For $\al
> 1$, the scaling law will thus be given by Eq. \ref{defect3}
(Eq. \ref{defect5}) for $\tau \ll(\gg) \tau_0$. We also note here
that our results do not apply to quenches which take a system
through a critical line \cite{sen2,pell1}.

\begin{figure}
\rotatebox{0}{\includegraphics*[width=\linewidth]{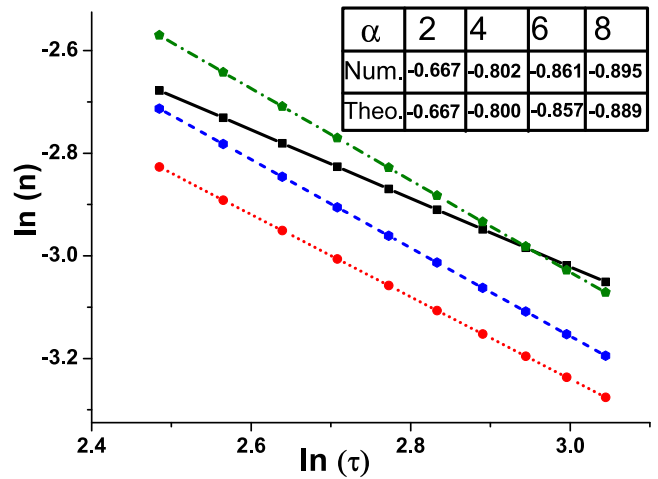}}
\caption{Plot of $\ln(n)$ vs $\ln(\tau)$ for the 1D Kitaev model for
$\al=2$ (black solid line), $\al=4$ (red dotted line), $\al=6$ (blue
dashed line) and $\al=8$ (green dash-dotted line). The slopes of
these lines agree reasonably with the predicted theoretical values
$-\al/(\al+1)$ as shown in the table.} \label{fig2} \end{figure}

We now supplement these analytical results with numerical studies of
well-known models. The first model that we choose for this purpose
is the one-dimensional Ising model in a transverse field with the
Hamiltonian $H_{\rm Ising} = -J (\sum_{\la ij\ra} S_i^z S_j^z -g_0
\sum_i S_i^x)$ where $J$ is the nearest neighbor coupling 
and $g_0$ is the dimensionless transverse field. A
standard Jordan-Wigner transformation \cite{sachdev1} then maps
$H_{\rm Ising}$ to a free fermionic Hamiltonian $H'_{\rm Ising}/J =
\sum_k \psi_k^{\dagger}[(g_0-\cos(k))\tau_3 + \sin(k) \tau_1]
\psi_k$. Thus a time variation $g_0(t)= |t/\tau|^{\al} {\rm
sign}(t)$ takes the system through two critical points at
$t_0=\tau(-\tau)$ where the energy gap vanishes at $k_0=0(\pi)$ so
that $g_0=1(-1)$ at these points. Thus the defect production around
both these critical points have the same $\tau_{\rm eff}(\al) =
\tau/\al$ and we expect (Eq. (\ref{defect4})) the defect density to
go as $n \sim \sqrt{\al}$ for a fixed $\tau$. To confirm this
expectation, we solve the time-dependent Schr\"odinger equation $i
\partial_t \psi(k,t) =H'_{\rm Ising} \psi(k,t)$ and compute the
defect probability $p_k$ and hence $n$ for fixed $\tau$ and for
several representative values of $\al \ge 1$. These values of $\al$
and $\tau$ are chosen so that we are in the regime where all
$R_{n>1}$ can be safely neglected. The plot of $n$ as a function of
$\al$ for $\tau=10,15,\,{\rm and}\,20$ is shown in Fig. \ref{fig1}.
A fit to these curves yields exponents of $0.506 \pm 0.006$
($\tau=10$), $0.504 \pm 0.004$ ($\tau=15$), and $0.505 \pm 0.002$
($\tau=20$) which are indeed remarkably close to the theoretical
value $1/2$ predicted by Eq. (\ref{defect4}). The systematic positive 
deviations in the exponents comes from the neglected terms $R_{n>1}$. 
We note that the range of $\al$ for which such deviation remains small 
grows with $\tau$, as expected from our theoretical prediction.

Next, we consider the one-dimensional Kitaev model
\cite{kit1,feng1,sen2} which has the Hamiltonian $H_K =\sum_{i \in
{\rm even}} \left( J_1 S_i^x S_{i+1}^x + J_2 S_i^y
S_{i-1}^y\right)$, where the sum extends over even sites $i$ on the
disconnected chains of the underlying hexagonal lattice, and ${\bf
S}_i$ denotes the spin at site $i$. Such a model can be realized as
the $J_3=0$ limit of the well-known Kitaev model and can be mapped,
via a standard Jordan-Wigner transformation \cite{feng1,sen2}, onto
the fermionic Hamiltonian $H'_K = 2 \sum_k \psi_k^{\da} \left( -
J_- \sin(k) \tau_3 + J_+ \cos(k) \tau_2\right) \psi_k = 2
\sum_k \psi_k^{\da} H'(k) \psi_k$ where
$\psi(k)=(c_1(k),c_2(k))$ are fermionic operators, $0\le k\le \pi$
extends over half the Brillouin zone, $J_{\pm} = J_1 \pm J_2$, and
we have chosen the lattice spacing to be unity. Here the time
variation $J_- (t) = J |t/\tau|^{\al} {\rm sign}(t)$, keeping
$J_+$ fixed, takes the system through a single critical point at
$t=0$ and $k=k_0=\pi/2$ which has $\nu=z=1$. The defect density,
according to Eq. (\ref{defect3}), is therefore expected to scale as
$n \sim \tau^{-\al/(\al+1)}$. To check this prediction, we
numerically solve the Schrodinger equation $i \partial_t \psi(k) =
H'_K (k;t) \psi(k,t)$ and compute the defect density $n
= \int_0^{\pi} dk/\pi \, p(k)$ as a function of the quench rate
$\tau$ for $\al$ with fixed $J_+ /J=1$. A plot of $\ln (n)$ as a
function of $\ln (\tau)$ for different values of $\al$ is shown in
Fig. \ref{fig2}. The slope of these lines, as can be seen from
Fig. \ref{fig2}, changes from $-0.67$ towards $-1$ as $\al$
increases from $2$ towards larger values. This behavior is
consistent with the prediction of Eq. (\ref{defect3}). The slopes of
these lines also show excellent agreement with Eq. (\ref{defect3})
as shown in the inset of Fig. \ref{fig2}.

Experimental verification of our results may be achieved in several
possible ways. First, there has been a concrete proposal for the
realization of the Kitaev model using an optical lattice\cite{duan}. In
such a realization, all the couplings can be independently tuned
using separate microwave radiations. In the proposed experiment, one
needs to keep $J_3=0$ and vary $J_{1(2)} = J(1\pm |t/\tau|^{\al}
{\rm sign}(t))/2$ so that $J_+$ remains constant while $J_-$
varies in time. The variation of the defect density, which in the
experimental setup would correspond to the bosons being in the wrong
spin state, would then show the theoretically predicted power-law
behavior (Eq. (\ref{defect3})). Secondly, a similar quench experiment
can be carried out with spin one bosons in a magnetic field
described by an effective Hamiltonian $H_{\rm eff} = c_2 n_0 \la {\bf
S} \ra^2 + c_1 B^2 \la S_z^2 \ra$ \cite{sk1} where $c_2 < 0$ and $n_0$
is the boson density. Such a system undergoes a quantum phase
transition from the ferromagnetic to polar condensate at $B^{\ast} =
\sqrt{|c_2|n_0/c_1}$. A quench of the magnetic field $B^2=B_0^2
|t/\tau|^{\al}$ thus would lead to scaling of defect density with an 
effective rate $\tau_{\rm eff} (\al)= \tau/(\al g^{(\al-1)/\al})$, 
where $g=|c_2|n_0 /c_1$. A measurement of
the dependence of the defect density $n$ on $\al$ should
therefore serve as test of prediction of Eq. (\ref{defect5}).
Finally, spin gap dimer compounds such as ${\rm Ba Cu Si_2 O_6}$ are
known to undergo a singlet-triplet quantum phase transition at $B_c
\simeq 23.5 $T which is known to be very well described by the mean-field
exponents $z=2$ and $\nu=2/3$ \cite{sebas1}. Thus a non-linear
quench of the magnetic field through its critical value $B=B_c + B_0
|t/\tau|^{\al}{\rm sign}(t)$ should lead to scaling of defects $n
\sim \tau^{-6\al/(4\al+3)}$ in $d=3$. In the experiment,
the defect density would correspond to residual singlets in the
final state which can be computed by measuring the total
magnetization of the system immediately after the quench.

To conclude, we have obtained general scaling laws of the defect density for 
an arbitrary power-law quench through a QCP which reproduce their linear 
counterpart as a special case. We have verified our theoretical prediction by 
numerical simulation of model systems and have suggested several possible 
experiments to test our results. Our results have been recently used to find 
the optimal passage through a QCP \cite{baran1}.

\end{document}